\documentclass[runningheads]{llncs}
\usepackage[T1]{fontenc}
\usepackage{graphicx}
\usepackage[backref=page]{hyperref}
\usepackage{color}

\usepackage{orcidlink}
\usepackage{xcolor}

\hypersetup{
    colorlinks=true,
    linkcolor=blue,
    urlcolor=blue,
    citecolor=black,
    filecolor=black,
    linkbordercolor=white, % Set the color of the hyperlink boxes to white
    urlbordercolor=white,
    citebordercolor=white,
    filebordercolor=white,
}

\usepackage{hhline}
\usepackage{multirow}
\usepackage{makecell}
\usepackage{comment}

\usepackage{siunitx}

\usepackage{float} %for figures
\usepackage[noabbrev]{cleveref}

\usepackage{color}

\usepackage[misc]{ifsym}

\begin{document}

\title{Anatomy-informed Data Augmentation\\for Enhanced Prostate Cancer Detection}
\titlerunning{Anatomy-informed Data Augmentation}
\author{
    Balint Kovacs\inst{1,2,3,\text{\Letter}}\orcidlink{0000-0002-1191-0646}\and
    Nils Netzer\inst{2,3}\orcidlink{0000-0001-5805-0535}\and
    Michael Baumgartner\inst{1,4,5}\orcidlink{0000-0003-4455-9917}\and
    Carolin Eith\inst{2,3}\and
    Dimitrios Bounias\inst{1,3}\and
    Clara Meinzer\inst{2}\and
    Paul F. Jäger\inst{5,6}\orcidlink{0000-0002-6243-2568}\and
    Kevin S. Zhang\inst{2}\and
    Ralf Floca\inst{1}\and
    Adrian Schrader\inst{2,3}\orcidlink{0000-0003-2112-0928}\and
    Fabian Isensee\inst{1,5}\orcidlink{0000-0002-3519-5886}\and
    Regula Gnirs\inst{2}\and
    Magdalena Görtz\inst{7,8}\and
    Viktoria Schütz\inst{7}\and
    Albrecht Stenzinger\inst{9}\and
    Markus Hohenfellner\inst{7}\and
    Heinz-Peter Schlemmer\inst{2}\and
    Ivo Wolf\inst{10}\orcidlink{0000-0002-6519-6484}\and
    David Bonekamp\inst{2,\star}\orcidlink{0000-0002-4811-0087}\and
    Klaus H. Maier-Hein\inst{1,11,\star}\orcidlink{0000-0002-6626-2463}
}
%index{Kovacs, Balint}
%index{Netzer, Nils}
%index{Baumgartner, Michael}
%index{Eith, Carolin}
%index{Bounias, Dimitrios}
%index{Meinzer, Clara}
%index{Jäger, Paul F.}
%index{Zhang, Kevin S.}
%index{Floca, Ralf}
%index{Schrader, Adrian}
%index{Isensee, Fabian}
%index{Gnirs, Regula}
%index{Görtz, Magdalena}
%index{Schütz, Viktoria}
%index{Stenzinger, Albrecht}
%index{Hohenfellner, Markus}
%index{Schlemmer, Heinz-Peter}
%index{Wolf, Ivo}
%index{Bonekamp, David}
%index{Maier-Hein, Klaus H.}

\authorrunning{B. Kovacs et al.}

\institute{
Division of Medical Image Computing,\\German Cancer Research Center (DKFZ), Heidelberg, Germany
\\\email{balint.kovacs@dkfz-heidelberg.de}\and
Division of Radiology, DKFZ, Heidelberg, Germany\and
Medical Faculty Heidelberg, Heidelberg University, Heidelberg, Germany\and
Faculty of Mathematics and Computer Science, Heidelberg University, Germany\and
Helmholtz Imaging, DKFZ, Heidelberg, Germany\and
Interactive Machine Learning Group, DKFZ, Heidelberg, Germany\and
Department of Urology, University of Heidelberg Medical Center, Germany\and
Junior Clinical Cooperation Unit 'Multiparametric methods\\for early detection of prostate cancer', DKFZ, Heidelberg, Germany\and
Institute of Pathology, University of Heidelberg Medical Center, Germany\and
Mannheim University of Applied Sciences, Mannheim, Germany\and
Pattern Analysis and Learning Group, Department of Radiation Oncology, Heidelberg University Hospital, Heidelberg, Germany
}

\maketitle
\footnotetext[1]{Equal contribution.}

\begin{abstract}
Data augmentation (DA) is a key factor in medical image analysis, such as in prostate cancer (PCa) detection on magnetic resonance images. State-of-the-art computer-aided diagnosis systems still rely on simplistic spatial transformations to preserve the pathological label post transformation. However, such augmentations do not substantially increase the organ as well as tumor shape variability in the training set, limiting the model's ability to generalize to unseen cases with more diverse localized soft-tissue deformations. We propose a new anatomy-informed transformation that leverages information from adjacent organs to simulate typical physiological deformations of the prostate and generates unique lesion shapes without altering their label. Due to its lightweight computational requirements, it can be easily integrated into common DA frameworks. We demonstrate the effectiveness of our augmentation on a dataset of 774 biopsy-confirmed examinations, by evaluating a state-of-the-art method for PCa detection with different augmentation settings.
\end{abstract}

\section{Introduction}
\label{sec:Introduction}
\par Data augmentation (DA) is a key factor in the success of deep neural networks (DNN) as it artificially enlarges the training set to increase their generalization ability as well as robustness \cite{shorten2019survey}. It plays a crucial role in medical image analysis \cite{isensee2021nnu} where annotated datasets are only available with limited size. DNNs have already successfully supported radiologists in the interpretation of magnetic resonance images (MRI) for prostate cancer (PCa) diagnosis \cite{bhattacharya2022review}. However, the DA scheme received less attention, despite its potential to leverage the data characteristic and address overfitting as the root of generalization problems.
\par State-of-the-art approaches still rely on simplistic spatial transformations, like translation, rotation, cropping, and scaling by globally augmenting the MRI sequences \cite{netzer2021fully,saha2021end}. They exclude random elastic deformations, which can change the lesion outline but might alter the underlying label and thus produce counterproductive examples for training \cite{shorten2019survey}. However, soft tissue deformations, which are currently missing from the DA schemes, are known to significantly affect the image morphology and therefore play a critical role in accurate diagnosis \cite{engels2020multiparametric}.
\par Both lesion and prostate shape geometrical appearance influence the clinical assessment of Prostate Imaging-Reporting and Data System (PI-RADS) \cite{weinreb2016pi}. The prostate constantly undergoes soft tissue deformation dependent on muscle contractions, respiration, and more importantly variable filling of the adjacent organs, namely the bladder and the rectum. Among these sources, the rectum has the largest influence on the prostate and lesion shape variability due to its large motion \cite{boubaker2017bladder} and the fact that the majority of the lesions are located in the adjacent peripheral prostate zone \cite{ali2022prostate}. However, only one snapshot of all these functional states is captured within each MRI examination, and almost never will be exactly the same on any repeat or subsequent examination. Ignoring these deformations in the DA scheme can potentially limit model performance.
\par Model-driven transformations attempting to simulate organ functions - like respiration, urinary excretion, cardiovascular- and digestion mechanics - offer a high degree of diversity while also providing realistic transformations. Currently, the finite element method (FEM) is the standard for modeling biomechanics \cite{payan2017biomechanics}. However, their computation is overly complex \cite{khallaghi2015statistical} and therefore does not scale to on-the-fly DA \cite{hu2018adversarial}. Recent motion models rely on DNNs using either a FEM model \cite{pfeiffer2019learning} or complex training with population-based models \cite{romaguera2021probabilistic}. Motion models have not been integrated into any deep learning framework as an online data augmentation yet, thereby leaving the high potential of inducing application-specific knowledge into the training procedure unexploited.
\par In this work we propose an anatomy-informed spatial augmentation, which leverages information from adjacent organs to mimic typical deformations of the prostate. Due to its lightweight computational requirements, it can be easily integrated into common DA frameworks. This technique allows us to simulate different physiological states during the training and enrich our dataset with a wider range of organ and lesion shapes. Inducing this kind of soft tissue deformation ultimately led to improved model performance in patient- and lesion-level PCa detection on an independent test set.

\begin{figure}[h] {}
    \begin{center}
        \includegraphics[width=0.99\textwidth]{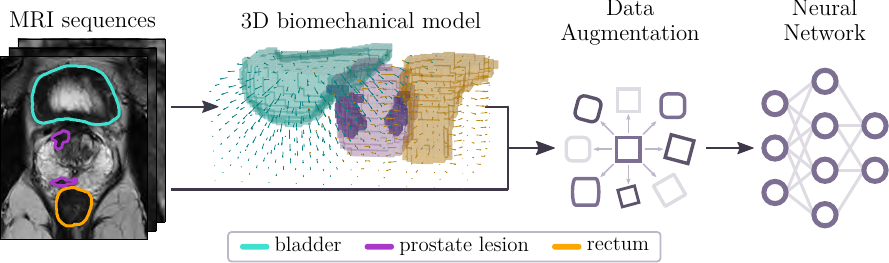}{}
        \caption{The proposed anatomy-informed prostate augmentation. Simulating typical physiologic changes in adjacent organs enlarges the training set with realistic soft tissue deformations of the prostate thereby increasing the generalization ability as well as the robustness of the network. Due to its lightweight computational requirements, it can be easily integrated into online network training. 
        }
        \label{fig:concept} {}
    \end{center} {}
\end{figure}

\section{Methods}
\subsection{\textbf{Mathematical model of the anatomy-informed deformation}}
    \label{subsec:semielx}
Model-driven spatial transformations simulate realistic soft-tissue deformations, which are part of the physiology, and can highly affect the shape of the prostate as well as the lesions in it. As the computation of state-of-the-art FEM models does not scale to on-the-fly DA, we introduce simplifications to be able to integrate such a biomechanical model as an online DA into the model training:
    \begin{itemize}
        \item soft tissue deformation of the prostate is mostly the result of morphological changes in the bladder and rectal space \cite{engels2020multiparametric,boubaker2017bladder},
        \item due to the isotropic mechanical behavior of the rectum and the bladder \cite{rubod2012biomechanical}, we apply isotropic deformation to them,
        \item we assume similar elastic modulus between the prostate and surrounding muscles \cite{qasim2022biomechanical}, allowing us to approximate these tissue classes as homogeneous,
        \item we introduce a non-linear component into the model by transforming the surrounding tissue proportionally to the distance from the rectum and bladder in order to generate realistic deformations \cite{wang2016patient}.
    \end{itemize}
Based on them, we define the vector field $V$ for the transformation as the gradient of the convolution between the Gaussian kernel $G_{\sigma}$ and the indicator function $S_{organ}$, multiplied by a scalar $C$ to control deformation amplitude and direction:
    \begin{equation}
        V = \nabla(G_{\sigma}*S_{organ}(x,y,z)) \cdot C.
    \end{equation}
The resulting $V$ serves as the deformation field for an MRI sequence $I(x,y,z)$:
    \begin{equation}
        I_{deformed}(x,y,z) = I(x + V_{x}(x,y,z), y + V_{y}(x,y,z), z + V_{z}(x,y,z)).
    \end{equation}
It allows us to simulate the distension or evacuation of the bladder or rectal space. We refer to this transformation as anatomy-informed deformation. We make it publicly available in Batchgenerators \cite{anonymousDAframework} and integrate it into a nnU-Net trainer \url{https://github.com/MIC-DKFZ/anatomy_informed_DA}.

\subsection{\textbf{Experimental setting}}
    \label{subsec:evaluation}
We evaluate our anatomy-informed DA qualitatively as well as quantitatively.
\par First, we visually inspect whether our assumptions in \Cref{subsec:semielx} regarding pelvic biomechanics resulted in realistic transformations. We apply either our proposed transformation to the rectum or the bladder, random deformable or no transformation in randomly selected exams and conduct a strict Turing test with clinicians having different levels of radiology expertise (a freshly graduated clinician (C.E.) and resident radiologists (C.M., K.S.Z.), 1.5 - 3 years of experience in prostate MRI) to determine if they can notice the artificial deformation.
\par Finally, we quantify the effect of our proposed transformation on the clinical task of patient-level PCa diagnosis and lesion-level PCa detection. We derive the diagnosis through semantic segmentation of the malignant lesions following previous studies \cite{duran2022prostattention,kohl2017adv,netzer2021fully,saha2021end,sanyal2020automated}. Semantic segmentation provides interpretable predictions that are sensitive to spatial transformations, making it appropriate for testing spatial DAs. To compare the performance of the trained models to radiologists, we calculate their performance using the clinical PI-RADS scores and histopathological ground truths.
To consider clinically informative results, we use the partial area under the Receiver Operating Characteristic (pAUROC) for patient-level evaluation with the sensitivity threshold of \SI{78.75}{\percent}, which is \SI{90}{\percent} of the sensitivity of radiologists for PI-RADS $\geq$ 4. Additionally, we calculate the $F_{1}$-score at the sensitivity of PI-RADS $\geq$ 4. Afterward, we evaluate model performances on object-level using the Free-Response Receiver Operating Characteristic (FROC) and the number of detections at the radiologists’ lesion level performance for PI-RADS $\geq$ 4, at 0.32 average number of False Positives per scan. Objects were derived by applying a threshold of $0.5$ to the softmax outputs followed by connected component analysis to identify connected regions in the segmentation maps. Predictions with an Intersection over Union of 0.1 with a ground truth object were considered True Positives. To systematically compare the effect of our proposed anatomy-informed DA with the commonly used settings, we create three main DA schemes:
\begin{enumerate}
    \item \textbf{Basic} DA setting of nnU-Net \cite{isensee2021nnu}, which is an extensive augmentation pipeline containing simple spatial transformations, namely translation, rotation and scaling. This setting is our reference DA scheme.
    \item \textbf{Random deformable} transformations as implemented in the nnU-Net \cite{isensee2021nnu} DA pipeline extending the basic DA scheme (1) to test its presence in the medical domain. Our hypothesis is that it will produce counterproductive examples, resulting in inferior performance compared to our proposed DA.
    \item Proposed \textbf{anatomy-informed} transformation in addition to the simple DA scheme (1). We define two variants of it:
        \begin{enumerate}
            \item Deforming only the rectum, as rectal distension has the highest influence among the organs on the shapes of the prostate lesions \cite{boubaker2017bladder}.
            \item Deforming the bladder in addition to the rectum, as bladder deformations also have an influence on lesions, although smaller.
        \end{enumerate}
\end{enumerate}

\subsection{\textbf{Prostate MRI Data}}
    \label{subsec:data}
\par 774 consecutive bi-parametric prostate MRI examinations are included in this study, which were acquired in-house during the clinical routine. The ethics committee of the Medical Faculty Heidelberg approved the study (S-164/2019) and waived informed consent to enable analysis of a consecutive cohort. All experiments were performed in accordance with the declaration of Helsinki \cite{world201864th} and relevant data privacy regulations. For every exam, PI-RADS v2 \cite{weinreb2016pi} interpretation was performed by a board-certified radiologist. Every patient underwent extended systematic and targeted MRI trans-rectal ultrasound-fusion transperineal biopsy. Malignancy of the segmented lesions was determined from a systematic-enhanced lesion ground-truth histopathological assessment, which has demonstrated reliable ground-truth assessment with sensitivity comparable to radical prostatectomy \cite{radtke2016multiparametric}. The samples were evaluated according to the International Society of Urological Pathology (ISUP) standards under the supervision of a dedicated uropathologist. Clinically significant prostate cancer (csPCa) was defined as ISUP grade 2 or higher. Based on the biopsy results, every csPCa lesion was segmented on the T2-weighted sequences retrospectively by multiple in-house investigators under the supervision of a board-certified radiologist. In addition to the lesions, the rectum and the bladder segmentations were automatically predicted by a model built upon nnU-Net \cite{isensee2021nnu} trained iteratively on an in-house cohort initially containing a small portion of our cohort. Multiple radiologists confirmed the quality of the predicted segmentations.

 \subsection{\textbf{Training protocol}}
    \label{subsec:training}
\par 774 exams were split into \SI{80}{\percent} training set (619 exams) and \SI{20}{\percent} test set (155 exams) by stratifying them based on the prevalence of csPCa (\SI{36.3}{\percent}). The MRI sequences were registered using B-spline transformation based on mutual information to match the ground-truth segmentations across all modalities \cite{netzer2021fully,pellicer2021deep}. As the limited number of exams with csPCa and the small lesion size compared to the whole image can cause instability during training, we adapted the cropping strategy from \cite{sanyal2020automated} by keeping the organ segmentations to use the anatomy-informed DA (offsets of $\pm$\SI{9}{\mm} axial to the prostate and $\pm$\SI{11.25}{\mm} in the axial plane to the rectum and the bladder). The images are preprocessed by the automated algorithm of nnU-Net \cite{isensee2021nnu}. We trained 3D nnU-Net models in 5-fold cross-validation with different spatial DA schemes, see \cref{subsec:evaluation}. The hyperparameter $C$ of the anatomy-informed DA was optimized using validation results, sampled during training with uniform distribution constrained by amplitude values in positive and negative directions of $C=\left\{300, 600, 900, 1200, 1500\right\}$. $C_{rectum}=1200$ and $C_{bladder}=600$ were selected for the final models. Compared to the standard nnU-Net settings, we implemented balanced sampling regarding the prevalence of csPCa and reduced the number of epochs to 350 to avoid overfitting. We used Mish activation function, Ranger optimizer, cosine anneal learning rate scheduler, and initial learning rate of $0.001$ following \cite{netzer2021fully}. The final models are ensembled and evaluated on the independent test set using bootstrapping with 1000 replications to provide standard deviation and to calculate p-values for the  $F_{1}$-score and for the number of detected lesions using two-sided t-test to determine statistical significance.

\section{Results}
    \label{sec:Results}
\begin{figure}[H] {}
		\begin{center}
			\includegraphics[width=0.71\textwidth]{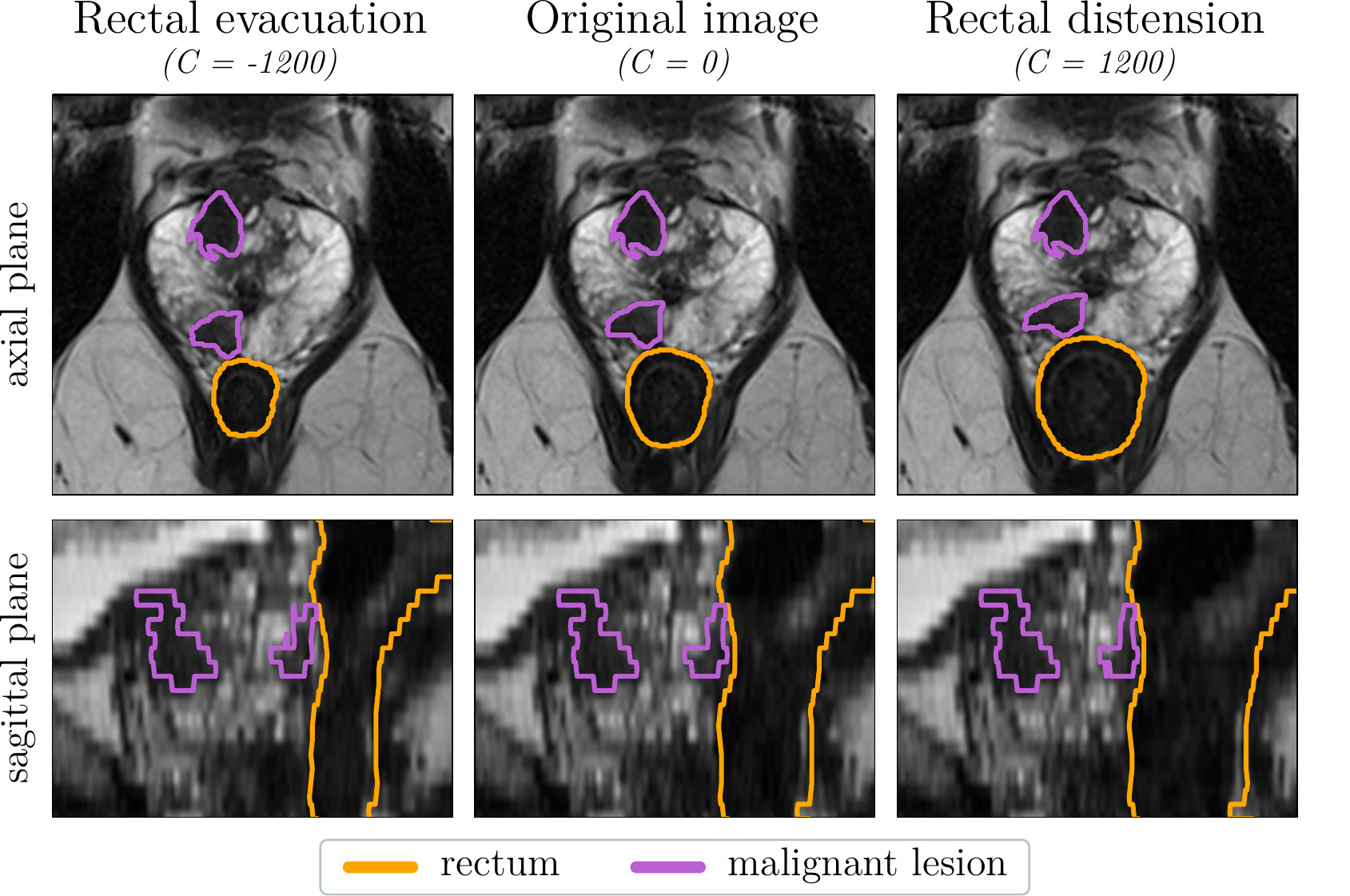}{}
			\caption{Results of the proposed anatomy-informed deformation on the rectum, showing the outlines of the rectum and malignant lesions. The middle images show the original MRI sequence, the left images simulate rectal space evacuation, while the right images rectal distension. The transformation induces localized soft tissue deformations, resulting in changes in the shape of lesions only in the adjacent peripheral prostate zone.}
			\label{fig:anatinf_transf} {}
		\end{center} {}
	\end{figure}

The anatomy-informed transformation produced highly realistic soft tissue deformations. \Cref{fig:anatinf_transf} shows an example of the transformation simulating rectum distensions with prostate lesions at different distances from the rectum. \SI{92}{\percent} of the rectum and \SI{93}{\percent} of the bladder deformation from the randomly picked exams became so realistic that our freshly graduated clinician did not detect them, but our residents noticed \SI{87.5}{\percent} of the rectum and \SI{25}{\percent} of the bladder deformations based on small transformation artifacts and their expert intuition. Irregularities resulted from the random elastic deformations can be easily detected, in contrast to our method being challenging to detect its artificial nature.
\par
\begin{figure}[H] {}
		\begin{center}
			\includegraphics[width=0.7\textwidth]{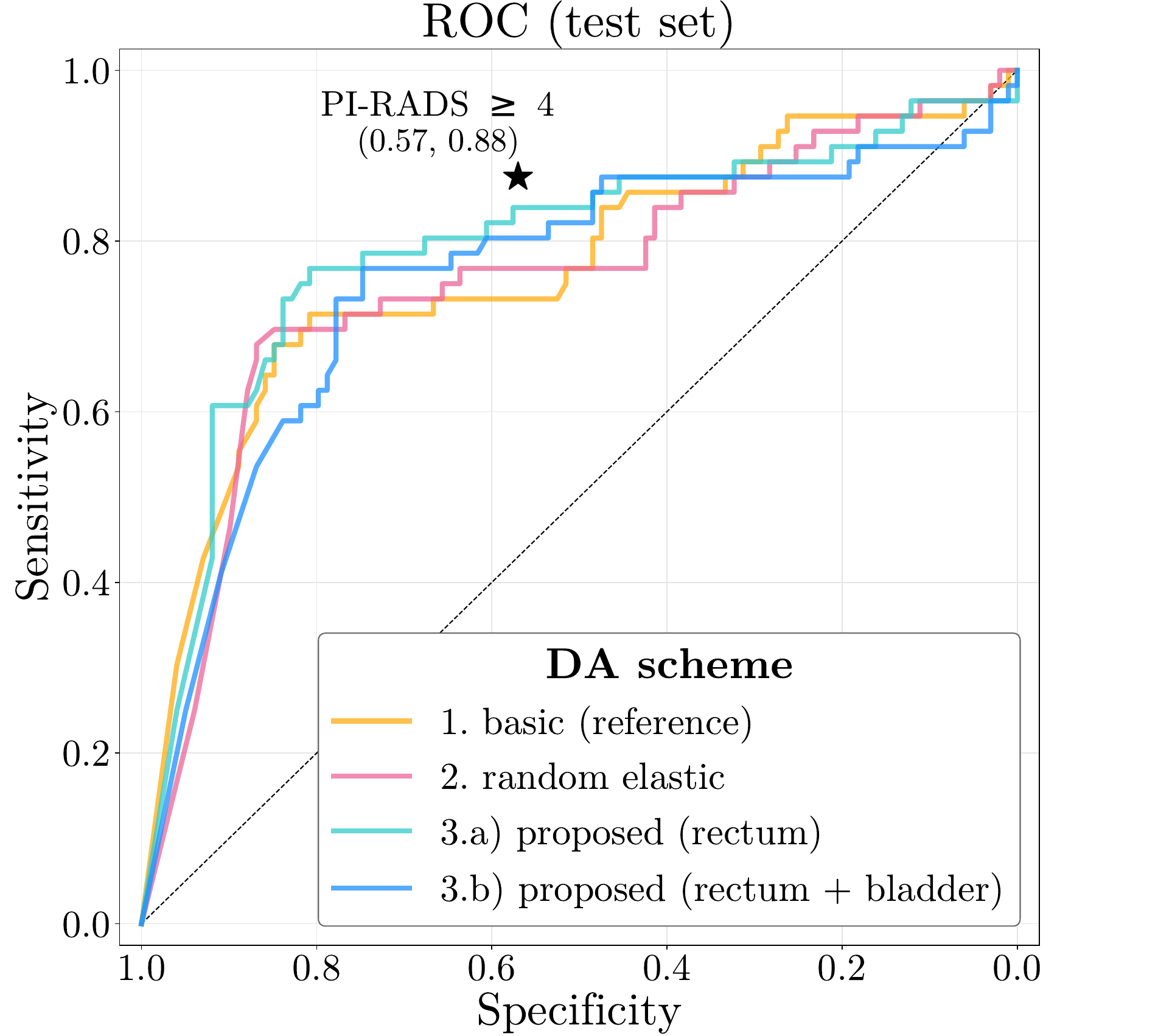}{}
			\caption{The effect of different spatial DA schemes on the ROC. The radiologists' performance with PI-RADS $\geq$ 4 is marked to locate the most informative performance point clinically. Both variants of the proposed anatomy-informed DA (3.a and 3.b) increased the sensitivity value around the clinical PI-RADS $\geq$ 4 performance point compared to the simple (1) and random elastic (3) DA schemes, approaching it closely.
            }
			\label{fig:results_ROC} {}
		\end{center} {}
	\end{figure}

 In \Cref{table:results} we summarize the patient-level pAUROC and $F_{1}$-scores; and lesion-level FROC results on the independent test set showing the advantage of using anatomy-informed DA. To further highlight the practical advantage of the proposed augmentation, we compare the performance of the trained models to the radiologists' diagnostic performance for PI-RADS $\geq$ 4, which locate the most informative performance point clinically on the ROC diagram, see \Cref{fig:results_ROC}.
    \begin{table}[H]
        \begin{center}
            %\footnotesize
        \caption{Prostate cancer detection results on our independent test set}
        \label{table:results}
            \begin{tabular}{p{0.395\textwidth}<{\raggedright}
            >{\raggedleft}p{0.09\textwidth}@{$\pm$}p{0.09\textwidth}<{\raggedright} 
            >{\raggedleft}p{0.075\textwidth}@{$\pm$}p{0.075\textwidth}<{\raggedright} 
            >{\raggedleft}p{0.075\textwidth}@{$\pm$}p{0.075\textwidth}<{\raggedright}}
        %\rowcolor{lightgray}
        DA scheme & \multicolumn{2}{c}{pAUROC} & \multicolumn{2}{c}{$F_{1}$-score} & \multicolumn{2}{c}{FROC}  \\ \hline
        1. basic (reference) &
        44.33 & \SI{11.65}{\percent} & 57.31 & \SI{3.14}{\percent} & 58.14 & \SI{5.79}{\percent} \\  
        2. random elastic &
        38.94 & \SI{14.38}{\percent} & 56.98 & \SI{3.08}{\percent} & 58.63 & \SI{5.42}{\percent} \\
        3.a) proposed (rectum) &
        59.92 & \SI{13.27}{\percent} & 61.64 & \SI{3.61}{\percent} & 59.55 & \SI{5.97}{\percent} \\
        3.b) proposed (rectum $+$ bladder) &
        53.27 & \SI{13.42}{\percent} & 62.42 & \SI{3.84}{\percent} & 59.93 & \SI{5.53}{\percent} \\
        \end{tabular}
        \end{center}
    \end{table}

\par Extending the basic DA scheme with the proposed anatomy-informed deformation not only increased the sensitivity closely matching the radiologists' patient-level diagnostic performance but also improved the detection of PCa on a lesion level. Interestingly, while the use of random deformable transformation also improved lesion-level performance, it did not approach the diagnostic performance of the radiologists, unlike the anatomy-informed DA.
\par At the selected patient- and object-level working points, the model with the proposed rectum- and bladder-informed DA scheme reached the best results with significant improvements ($p<0.05$) compared to the model with the basic DA setting by increasing the $F_{1}$-score with \SI{5.11}{\percent} and identifying 4 more lesions (\SI{5.3}{\percent}) from the 76 lesions in our test set.
\par The time overhead introduced by anatomy-informed augmentation caused no increase in the training time, the GPU remained the main bottleneck.

\section{Discussion}
This paper addresses the utilization of anatomy-informed spatial transformations in the training procedure to increase lesion, prostate, and adjacent organ shape variability for the task of PCa diagnosis. For this purpose, a lightweight mathematical model is built for simulating organ-specific soft tissue deformations. The model is integrated into a well-known DA framework and used in model training for enhanced PCa detection.
\newline\textbf{Towards radiologists' performance} Inducing lesion shape variability via anatomy-informed augmentation to the training process improved the lesion detection performance and increased the sensitivity value towards radiologist-level performance in PCa diagnosis in contrast to the training with the basic DA setting. These soft tissue deformations are part of physiology, but only one snapshot is captured from the many possible functional states within each individual MR examination. Our proposed DA simulates examples of physiologic anatomical changes that may have occurred in each of the MRI training examples at the same exam time points, thereby aiding the generalization ability as well as the robustness of the network. We got additional, but slight improvements by extending the DA scheme with bladder distensions. A possible explanation for this result is that less than \SI{30}{\percent} of the lesions are located close to the bladder, and our dataset did not contain enough training examples for more improvements. 
\newline\textbf{Realistic modeling of organ deformation} Our proposed anatomy-informed transformation was designed to mimic real-world deformations in order to preserve essential image features. Most of the transformed sequences successfully passed the Turing test against a freshly graduated clinician with prostate MRI expertise, and some were even able to pass against radiology residents with more expertise. To support the importance of realism in DA quantitatively, we compared the performance of the basic and our anatomy-informed DA scheme with that of the random deformable transformation. The random deformable DA scheme generated high lesion shape variability, but it resulted in lower performance values. This could be due to the fact that it can also cause implausible or even harmful image warping, distorting important features, and producing counterproductive training examples. In comparison, our proposed anatomy-informed DA outperformed the basic and random deformable DA, demonstrating the significance of realistic transformations for achieving superior model performance.
\newline\textbf{High applicability with limitations} The easy integration into DA frameworks and no increase in the training time make our proposed anatomy-informed DA highly applicable. Its limitation is the need for additional organ segmentations, which requires additional effort from the annotator. However, pre-trained networks for segmenting anatomical structures like nnU-Net \cite{isensee2021nnu} have been introduced recently, which can help to overcome this limitation. Additionally, our transformation computation allows certain errors in the organ segmentations compared to applications where fully accurate segmentations are needed. The success of anatomy-informed DA opens the research question of whether it enhances performance across diverse datasets and model backbones.

\section{Conclusion}
In this work, we presented a realistic anatomy-informed augmentation, which mimics typical organ deformations in the pelvis. Inducing realistic soft-tissue deformations in the model training via this kind of organ-dependent transformation increased the diagnostic accuracy for PCa, closely approaching radiologist-level performance. Due to its simple and fast calculation, it can be easily integrated into DA frameworks and can be applied to any organ with similar distension properties. Due to these advantages, the shown improvements in the downstream task strongly motivate to utilize this model as a blueprint for other applications.

\bibliographystyle{splncs04}
\bibliography{arxiv}

\newpage
\appendix
\section{Supplementary Material}

\begin{table}[H]
        \begin{center}
        %\footnotesize
        \caption{Parameters required for the implementation of the anatomy-informed data augmentation}
        \label{table:transformparams}
        \begin{tabular}{p{0.30\textwidth}<{\raggedright}
        p{0.30\textwidth}<{\centering}}
    %\rowcolor{lightgray}
 
    Parameter & Value(s)  \\ \hline
    Image resolution & (0.3125, 0.3125, 3.0)mm\\
    Spacing ratio    & 0.3125/3\\
    $\sigma$         & 32\\
    $C_{rectum}$     & 1200\\
    $C_{bladder}$    & 600\\
    probability      & 0.2
    \\ \hline

        \end{tabular}
        \end{center}
\end{table}

\begin{table}[H]
\begin{center}
\caption{Statistical significance results from the two-sided t-test in the selected patient- and object-level working points}
\begin{tabular}{p{0.4\textwidth}<{\raggedright}
        p{0.14\textwidth}<{\centering}
        p{0.14\textwidth}<{\centering}
        p{0.14\textwidth}<{\centering}
        p{0.14\textwidth}<{\centering}}
 & \multicolumn{2}{c}{Patient-level} & \multicolumn{2}{c}{Object-level} \\
DA scheme & $F_{1}$-score & p-value & Detections & p-value \\ \hline
1. basic (reference)     & 57.31         &   -     & $39/76$          &     -   \\
2. random elastic       & 56.98         & 0.13    & $42/76$          & $<0.01$ \\
3.a) proposed (rectum)       & 61.64         & $<0.01$ & $43/76$          & $<0.01$ \\
3.b) proposed (rectum $+$ bladder)      & 62.42         & $<0.01$ & $43/76$          & $<0.01$ \\ \hline
\end{tabular}
\end{center}
\end{table}

\end{document}